# High-performance automatic categorization and attribution of inventory catalogs

Kolonin A.G.[1]

[1] *Webstructor project*

**Annotation.** *Techniques of machine learning for automatic text categorization are applied and adapted for the problem of inventory catalog data attribution, with different approaches explored and optimal solution addressing tradeoff between accuracy and performance is selected.*

## 1 Overview

The goal of the research was to develop technology capable to decompose raw textual descriptions of product inventory catalog items producing categories (such as UNSPSC product category) assigned to them and attribute values (such as purpose and type of an item, type, vendor, material, color, weight, size, etc.) set. The following requirements have been posed.

- Learn the rules for category assignment and attributions from the training corpus - series of raw text descriptions accompanied with categories assigned and attribute values determined.
- Be able a apply the learned rules to novel textual descriptions not experienced before – so that learned rules enable category assignment and extraction of the attribute values for the each of new objects created for the each of novel textual descriptions.
- Employ the generic-purpose statistical and machine learning techniques.
- Be capable to deal with gigabytes of data and millions of input data texts and output objects in real time with multiple items per second.
- Display quality of recognition comparable to those of humans.

Since the source of decomposition was unstructured item description text and the destination was structured set of object attribute values, the problem could be defined as **objective text mining**.

## 2 Implementation model

Conceptual and algorithmic framework of text mining with account to probabilistic inference has been inspired by [1] and [2], while architectural and practical considerations were considered with account to [3], [4] and [5].

The simplicity of implementation model has been dictated by the requirement for efficient (in terms of run-time performance) solution. In terms of entity-relationship model, it was based on limited set of nullary and unary relations (entities) interconnected with binary (many-to-many) relations. Some of the relations were actually maintained by the implementation system (as containing learned rules) and some of them were not stored persistently (most of all due to requirement of saving storage space and increasing run-time performance requirements) but were instantiated transiently only for the operations with particular instances of entities.

Each relation instance in our data model had an **evidence** measure attached to it. The exact meaning of the evidence varied from entity to entity. Theoretically, the evidence could be positive or negative but we considered only the positive evidence in this design. Below we describe the purposes of the entities having the evidence values defined.

- **Text** – the key data source - each text entity instance may fall under particular text type. Evidence equals 1, normally, but may be set as floating point value in range [0,1] to indicate an imperfect quality of the text for the decomposition.
- **Token** – a persistent vocabulary element created from the texts of training corpus. Each text of training corpus is parsed/tokenized and each token is stored in the vocabulary. Each token may have many-to-many transient relationships to the texts which are parsed to this token and also many-to-many persistent relationships to the features based on this token. Evidence equals 1.
- **TextToken** – a transient linkage entity serving many-to-many relationship representing occurrence of a token in the text. Evidence equals to number of the token occurrences in the text.
- **Feature** – an entity representing **linguistic feature**. The feature has persistent many-to-many relationships (via linkage entities) to the tokens it consists of and to the categories indicated by it. Evidence equals 1 or may be set as floating point value in range [0,1] to indicate an imperfect quality of the feature.
- **FeatureToken** – a persistent linkage entity serving many-to-many relationship representing usage of a token in the feature. Evidence equals 1.
- **TextFeature** – a transient linkage entity serving many-to-many relationship representing appearance of a feature in a text. Evidence equals to number of the feature appearances in the text.
- **CategoryFeature** – a persistent linkage entity serving many-to-many relationship representing appearance of a feature in a text. Evidence is a floating point number residing in range [0,1] and indicating the inferred (by system) relevance of the feature to the category.
- **Category** – the primary **carrier of the meaning** decomposed from the raw text. Each category belongs to particular domain. The category forms persistent many-to-many relationships to the features used to indicate the appearance of the category and to the objects referring to this category via the values of their attributes. Evidence equals 1, normally, but may be set as floating point value in range [0,1] to indicate uncertainly defined categories.
- **TextCategory** – a transient linkage entity serving many-to-many relationship representing relevance of a category to the text. Evidence equals to 1 for the training set couples of texts accompanied with appropriate decomposition objects. Evidence resides in range [0,1] for the novel texts with the inferred relevant categories.
- **Domain** – grouping of multiple categories carrying the same sort of meaning, may be referenced by many **attributes** of different **classes**. Evidence does not matter. In our case domains are groups of UNSPSC taxonomy codes, lists of object type and purpose and finally possible values two tens of item attributes.
- **Class** – set of the **objects** sharing the same set of **attributes**. Evidence does not matter.
- **Attribute** – property of the **class** indicating its instance **objects** may take **values** as **categories** under particular **domain**. Evidence does not matter. In our design, we assume equivalence of Domain and Attribute, so that Attribute is bound to particular Domain while Domain implies specific Attribute.
- **CategoryDomain** – a persistent linkage entity serving many-to-many relationship representing co-occurrence of a category with domains. Evidence equals to number of times the the category co-occurs with any category of specific domain. In our design it is implied categories such as UNSPSC code, purpose and type dive appearance of all of the remaining domain attributes.
- **Object** – **instance** of the class. Evidence does not matter.
- **Value** – category filling the attribute of particular class instance object. Evidence does not matter.

Among the relations, there were **instant** relations and **inferable** relations. Instant entities were created by the system unconditionally with the evidence value equal to 1 or greater and the innate true confirmation status (which is technically equivalent). Inferred entities could get created by the system in the course of training or recognition inference processes. For the **training process**, inferable instance were of CategoryFeature entity so we were learning which features were specific to which categories given the training corpus. For **recognition process**, inferable instances were of TextCategory entity, so we were guessing which categories in the text description are to be recognized given the scope of rules learned before.

The following section describes the process of learning rules for item text decomposition into categories as values of attributes specific to the objects of given class, assuming the rule is represented as set of features thought indicative for particular categories. Given the training corpus where texts are associated with manually defined categories for class attributes, it creates sets of features indicating those categories.

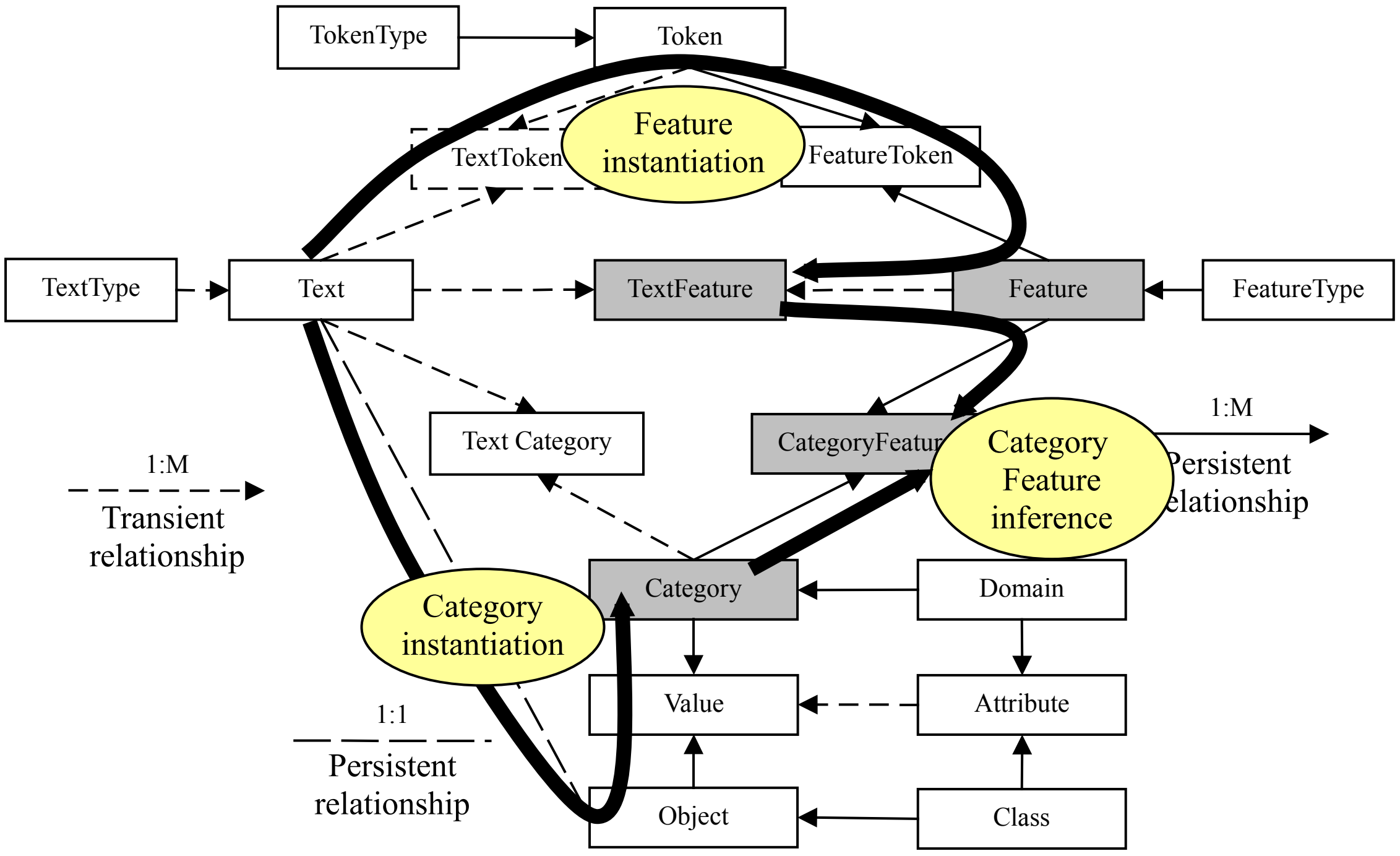

**Fig.1.** Entity-relationship model of the text classification framework and learning phase.

On the diagram above, there are three sub-processes contributing to the learning process.

- The first process is Category instantiation which takes the attributes defined for text in training corpus (either encoded in the text as tags or taken from respective database table fields) and creates categories for them, given the domain indicated by the attribute.
- The second process is Feature instantiation which takes the text in training corpus and decomposes it into tokens and features accordingly to the parser, tokenizer and feature builder depending on the implementation.
- The two processes above are independent, but they precede the third process which is Category Feature inference. It employs statistics to infer the relevance of features encountered in the texts to the categories associated with those texts.

The following section describes the process of applying the rules learned before for item text decomposition into categories as values of attributes specific to the objects of given class, assuming the rule is represented as set of features thought indicative for particular categories. Given the novel text and the class of the object, it uses the sets of learned features for categories under domains of the attributes specific to class to decide which categories represent appropriate values for those attributes.

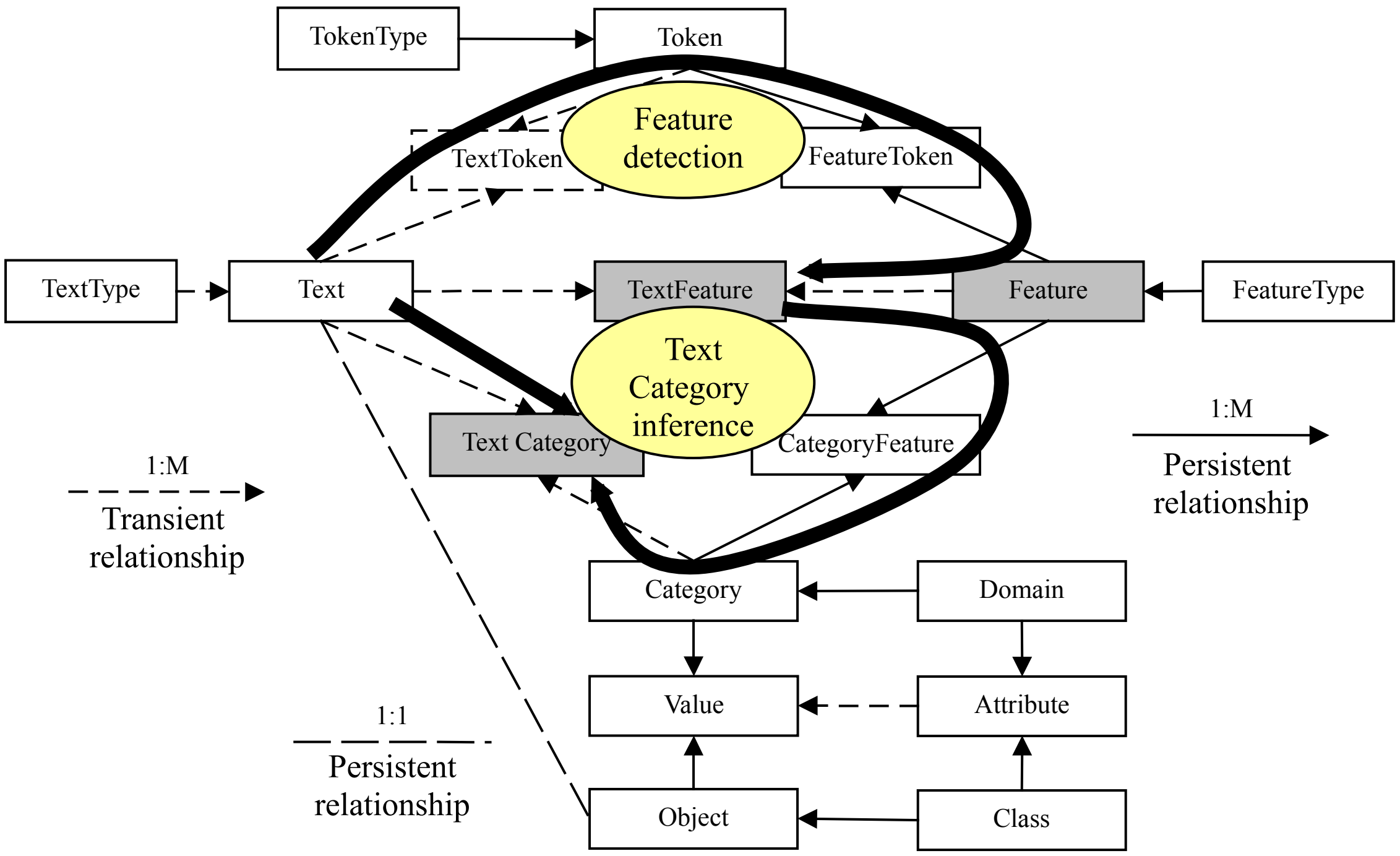

**Fig.2.** Entity-relationship model of the text classification framework and recognition phase.

On the diagram above, there are two sub-processes contributing to the recognition process.

- The first process is Feature detection which takes the text in novel data and decomposes it into tokens and features accordingly to the parser, tokenizer and feature builder depending on the implementation. This process is similar to Feature instantiation in the course of learning, but the key difference is that only the features instantiated earlier in the course of learning can be detected, no new features are instantiated.
- The second process is Text Category inference. It employs statistics to infer the relevance of texts to the categories associated with those texts through the features detected in the texts and learned for those categories.

We were using simple yet intuitively solid statistical model. The model was based on those entities in the data model discussed above that can have a meaningful evidence measure – as described on the following diagram.

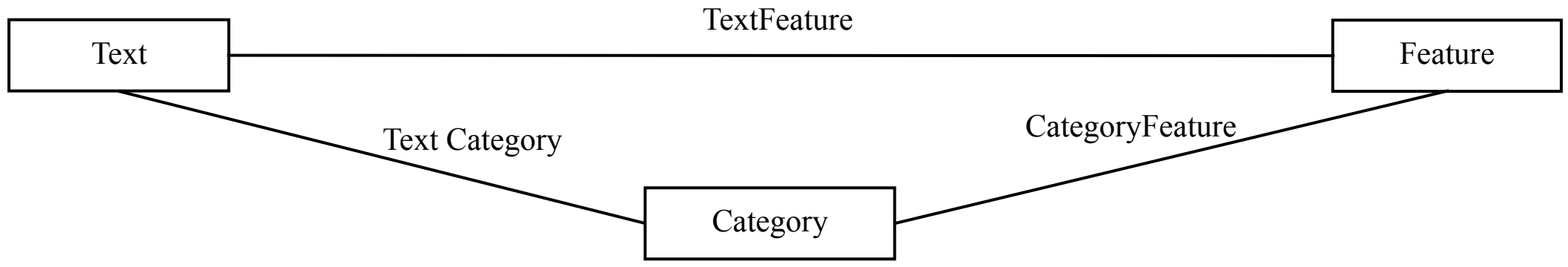

**Fig.2.** Basic model of relationships between texts, features and categories.

The following formulae provides ability to discover the significant features representing the text, rank the importance of categories associated with text and evaluate the representativity of features in respect to the categories. The formulae employ the probabilistic measure of mutual relevances of the experienced entities.

$TF_{tf}$ – **text feature evidence**, cadinal number counting occurences of feature f in text t ([0..N])
$F_t = \ _f TF_{tf}$ – features associated with text (**text feature evidence base**)

$T_f = \quad {}_t TF_{tf}$ – texts associated with feature (**feature text evidence base**)
T – number of texts
F – number of features
$TFTF_{tf} = TF_{tf}{}^2/(T_f * F_t)$ [0..1] – **mutual relevance** of the feature f and the text t.

The interpretation of the $TFTF_{tf}$ shows that there are two situations when its value can be high. First, when there are only few features in the text ($F_t$ is low), so that feature is highly contributive to the text, when compared to other features. Second, when there are only few texts are associated with feature ($T_f$ is low), so that feature is highly indicative in respect to the text, when compared to other texts. The extreme situation is that only one text contains given feature and it is the only one feature in the text ($TFTF_{tf}$=1).

Normally, the $TFTF_{tf}$ values are driven by the implementation of the parser/tokenizer and the set of the different features of various feature types defined in the implementation system. That is, they are not subject of manual confirmation – their confirmation is innate.

$TC_{dc}$ – **text category evidence**, cadinal number of times text t is though to contain a reference to category c ([0..N])
$C_t = \quad {}_c TC_{tc}$ – categories of given domain associated with text (**text category evidence base**)
$T_c = \quad {}_t TC_{tc}$ – texts associated with category (**category text evidence base**)
T – number of texts
C – number of categories under particular domain
$TCTC_{tc} = TC_{tc}{}^2/(T_c * C_t)$ [0..1] – **mutual relevance** of the category c and the text t.

The interpretation of the $TCTC_{tc}$ shows that there are two situations when its value can be high. First, when there are only few categories of the domain associated with the text ($C_t$ is low), so that category is highly contributive to the text, when compared to other categories. Second, when there are only few texts are associated with category ($T_c$ is low), so that category is highly indicative in respect to the text, when compared to other texts. The extreme situation is that only one text is associated with given category and it is the only one category that is associated with the text ($TCTC_{tc}$=1).

$CF_{cf}$ – **category feature evidence**, number of times a text with feature f is thought to fall under category c, scaled by multiplication with text-feature and text-category mutual relevance values ([0..1]). If the relation is confirmed manually, this becomes "all-or-none" ([0,+1]) value.
$F_c = \quad {}_f CF_{cf}$ – value of features associated with category (**category feature evidence base**)
$C_f = \quad {}_c CF_{cf}$ – value of categories associated with feature (**feature category evidence base**)
C – number of categories
F – number of features
$CFCF_f = CF_{cf} / C_f$ [0..1] – **feature category relevance** of the feature f and the category c.
$CFCF_f = CF_{cf} / F_c$ [0..1] – **category feature relevance** of the feature f and the category c.
$CFCF_{cf} = CF_{cf}{}^2/(C_f * F_c)$ [0..1] – **mutual relevance** of the feature f and the category c.

The interpretation of the $CFCF_{cf}$ is not as straightforward, as it is inferable value. However, for confirmed features ($CF_{cf}$ takes [0,+1]), there are two situations when its value can be high. First, when there are only few categories of the domain associated with the feature ($C_f$ is low), so that category is highly specific to the feature, when compared to other categories. Second, when there are only few features are associated with category ($F_c$ is low), so that feature is highly specific to the category, when compared to other texts. The extreme situation is that only one feature is associated with given category and it is the only one category that is associated with the feature ($CFCF_{cf}$=1).

# 3 Performance optimization and accuracy tuning

The problem we faced could be called as massive pattern recognition problem giving limited hardware/software (CPU and memory, primarily) resources, with the following issues.

1. The massive pattern recognition requires to access to most (if not all) of the data at once. Practically, it is unlikely possible that one can cache-out some portion of your data to swap (page) file on the disk and keep working with small partition of your data in RAM for a while.
2. The I/O response time for hard drive (swap/page file) is order of magnitude times higher than the I/O response time for RAM.
3. Per 1 and 2, the massive pattern recognition application which runs with some reasonable speed as long as its memory footprint fits available RAM, runs order of magnitude times slower if its memory footprint does not fit the RAM (and so all the time is spent on page file swapping).
4. The precision/accuracy/quality of the massive pattern recognition depends, beyond the algorithm, on amount and complexity of used information. The greater amount and complexity of information, the more memory is required.

For example, if one wants a recognition algorithm to run 2 times faster, one can reduce the best categories per feature from 10 to 5, but then will lose 10% of precision. For another example, if one wants recognition to be more accurate 10%, they need to involve ordered token series in analysis which will consume ten times more RAM. Time is memory and memory is time and both worth money. After all, the human brain deals with processing of huge amount of information in the same way – low reaction time typically imply lower accuracy of an action while precise action needs substantial amount of time preparing the reaction.

During initial experiments, it has been turned clear that in order to achieve target performance, no any disk operations (τeither database read/writes nor swapping/paging) have to be involved so entire implementation has been mad on top of Java arrays and-hash-table, given required amount of server memory allocated to the Java process.

Further, our goal was to fine-tune algorithm and implementation based on probabilistic statistical model with some additional operations and heuristics, enabling us to increase the performance and quality of category assignment and attribution to reasonable extent, given the tradeoff between the performance and quality. Many of these operations and heuristics have been tried but below is the short list of those that have turned successful.

## 3.1 Controlled feature instantiation

In brief, the parsing approach has been put under strict control with the number of restrictions. Since use of **keyword frame** features based on token combinations had turned to burst of total number of features (which apparently could not be handled by available operating memory), we realized a need for intelligent reduction of "token vocabulary" so only the usable tokens are used to build the **keyword** and **keyword frame** features, so there is a need to develop an intelligent algorithm for compression of the "feature vocabulary", including the following.

- For simplicity, we didn't consider **stem** features for now, so only distinguishable textual tokens were used, without of stemming them.
- As it had been made evident that feature tokens with numbers and shorter than 3 may be usable (such as abbreviations of 1 and 2 characters), we didn't have to discard them.
- As it had been proven the ordered token sets are useful to increase accuracy, we considered **keyword frame** features (as ordered sequences of keywords) but limited the number of tokens in the frame down to 2. To deal with text sequences like "surgery scissors" as well as "surgery steel scissors", we were building keyword frames hoping over the keywords, so that "office steel scissors" would effect in frames like "office-steel", "steel-scissors" and "office-scissors". When instantiating the features, the hoping distance was restricted to maxFrameDistance parameter (default 2). When parsing the features, the evidence of the frame feature would be set inverse to distance (so that "office scissors steel" would be parsed to "office-scissors" with evidence 1, "scissors-steel" with evidence 1 and "office-steel" with evidence 0.5).

## 3.2 Contextual scoping (restricting sets of Attributes)

Assuming a combintaion of categories such as UNSPSC category, purpose and type altogether essentially identifies a node on item taxonomy tree, we considered this as a context for narrowing the scope of things to consider. The purpose of this was to eliminate CPU time of spent of recognition of attributes typically not specific to an object type or category.

That is, when performing recognition for an item, we were recognizing only limited set of attributes, specific to given combination of values for UNSPSC, type and purpose attributes. To figure out the restricted subset, we could use intersection or union. For example, assuming we know the type=X1 is attributed with A5 and A6 (with variety of accompanying types) and purpose=Y1 is attributed with A6 and A7 (with variety of accompanying purposes), the following attributes were considered for combination type=X1, purpose=Y1:
- A5,A6,A7 - union of type and purpose associates
- A6 only - intersection of type and purpose associates

## 3.3 Restricting to best Feature Categories (Category-Feature space compression)

In order to gain decomposition performance in the course of recognition, sacrificing by quality, we decided to have an option to specify maximum number of best categories per feature considered. Maximum performance and lowest quality were granted by value of 1 while minimum performance and best quality were provided by value of corresponding to infinity. The reasonable value would be somewhere in between 10 and 100. Removing the extra category-feature relationships made the inferred association set more compact and effecting in better run-time performance. On the other hand, compression has turned to lose the quality inevitably.

## 3.4 Domain-specific specialization

The other sort of "compression" used was implementation of technique removing the rules for all domains other than domain of interest and not involving the domains other than domain of interest in recognition. This has been done in order to be able to recognize specific attribute only (e.g. UNSPSC code or purpose alone) for specific data set much more quickly. With fully "compressed rules" (specific to given domain), performance boosted dramatically up from tens/hundreds to thousand of items per minute – all because of the load relaxed in respect to data structures (hash-tables) involved in implementation.

## 3.5 Priority on order

When finding a suitable category given the list of features, we considered only the high-order features (**keyword frame** features, namely) first. If existed, we used them only and didn't consider low-order features (**keyword** features, namely) at all. If no high-order features existed, only then we considered the remaining low-order features.

## 3.6 Boolean ranking

When, given the list of categories created from all features, we were trying to rank the categories to figure out the best one, we were doing the following. Prior to using the inferred $TC_{tc}$ value, we were counting the number of features matched per each of the categories and giving higher rank to category which has more features matched. After then, we were breaking the ties with standard $TC_{tc}$ value. Practically, use of this approach generally improved recognition quality when dealing with "known" items (ones that had been encountered earlier in training sets) however it made recognition slightly worse when handling "unfamiliar" items (not ever seen in training sets before).

## 3.7 Evidence normalized by categories per feature

Instead of using the "symmetric" ("undirected") **mutual relevance** ($CFCF_{cf} = CF_{cf}^{\,2}/(C_f * F_c)$), we tried to use "asymmetric" ("directed") **feature-category relevance** ($CFCF_f = CF_{cf} / C_f$). This turned out to supply decent quality boost.

The latter turned to be slightly surprising (though understandable from cognitive perspective after all) that the original probabilistic formula of **mutual relevance** turned to be not that valuable for practical categorization inferences. Really, it worked well, emitting the matches around 65-75%. However, in order to get the decent accuracy on matching - around 75-99%, we just needed to use the much more simple (and faster in terms of run-time performance) asymmetric formula of $CFCF_f = CF_{cf} / C_f$.

That is, practically, if we are about to detect the category given the list of features, the account for associations of a given category with many features turns to be more than just meaningless, it turns to be misleading.

One might wonder if this held up if doing some "feature selection" and just use the most important features by some criterion. Indeed, this did not - as experiment had shown. An opposite had been experienced - any forms of "restricted feature selection" tried - all were bringing accuracy down. That underlined the original striking point - ranking features from category perspective does not play an interesting role when compared to ranking categories against features. Which means, if we are restricting features on the category-by-category basis, we just loose meaningful features.

# 4 Results and conclusions

The implementation of the model discussed above has been implemented in the course of proprietary project using Sun Java Development Kit and thoroughly tested using 32Gb and 64Gb Linux servers against real-world product inventory catalog data consisting of few millions of items in training and testing sets, few tens of millions of relation instances in total, half million category and attribute values with average few thousand values in each domain (up to two hundred). Given average 1K memory consumed by each relation this effected in memory footprint of few tens of gigabytes.

The most important result on quality of automatically categorized and attributed items appeared in the process of analyzing particular failures of the system to recognize and categorize some items properly. It has been found that one quarter of failures have been experienced due to errors made my humans creating training sets and so the wrong data has been used to infer wrong inference rules. The other quarter has turned to appear just because of similar errors made by humans creating testing data set. The other two quarters were due to insufficient data in training corpus (so that instantly unexperienced data had been tried to recognize) and finally to lack of software intelligence (due to restriction on categories per feature, missing true boolean regression and so forth) compared to human.

However, overall result turned that general intelligence of software system, in terms of recognition quality turned to be comparable (76-96%) to one of humans, providing much higher performance, as displayed on the following diagram.

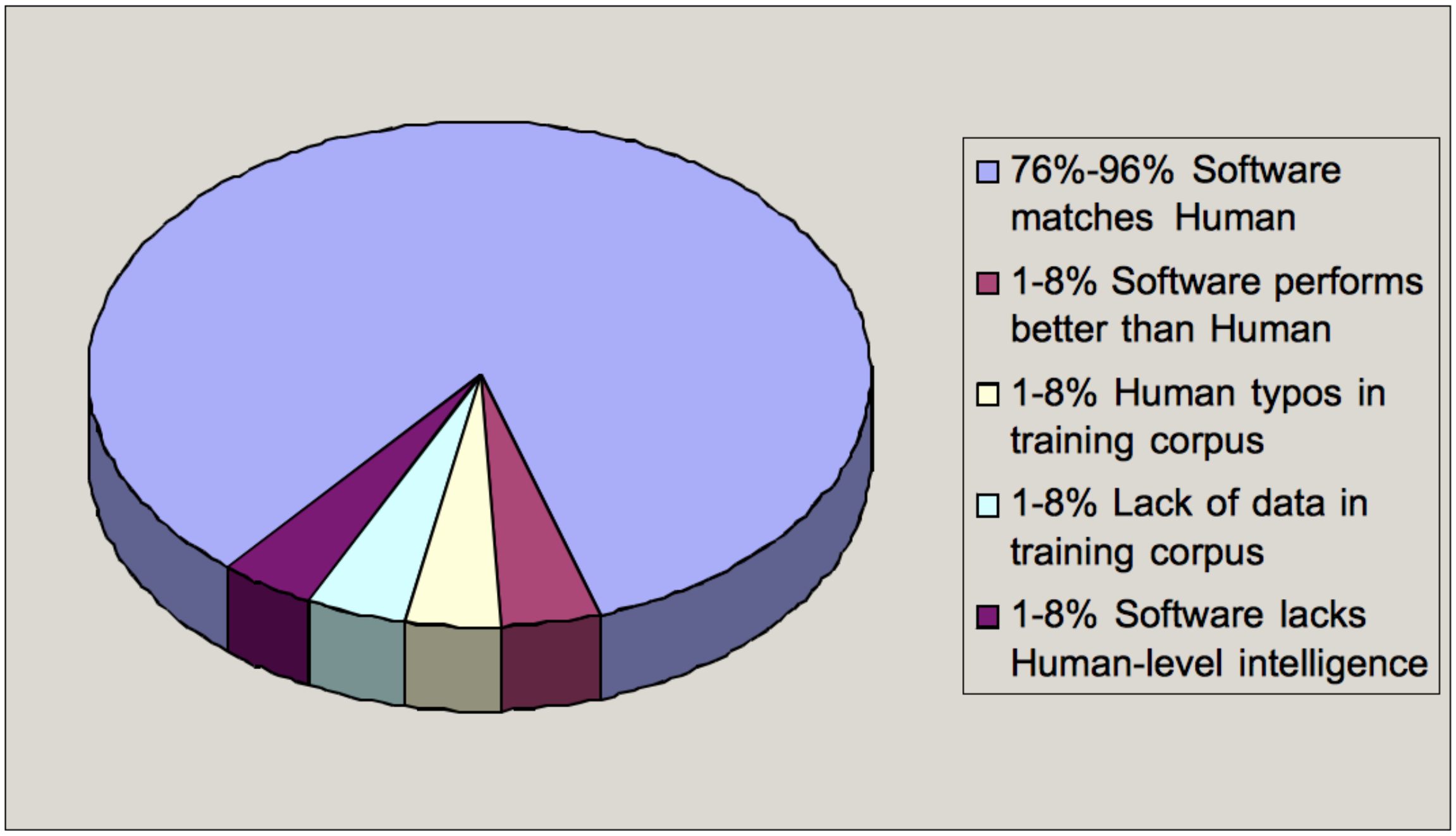

**Fig.2.** Diagram presenting comparative accuracy analysis.

At the same time, it has been shown that various optimization techniques, starting with architecture-level and code-level optimizations and ending with various heuristics and algorithm simplifications led to enormous increase of performance (few orders of magnitude - starting with more one second spent on single item ending with few hundred items per second) boost. Which is substantially outperforming human capabilities.

## Resources